\documentclass{elsart}

\usepackage[square,comma]{natbib}
\usepackage{graphicx}
\usepackage{amssymb}

\begin{document}

\thispagestyle{empty}
\begin{Large}
\textbf{DEUTSCHES ELEKTRONEN-SYNCHROTRON}

\textbf{\large{in der HELMHOLTZ-GEMEINSCHAFT}\\}
\end{Large}

DESY 05-137

August 2005

\begin{eqnarray}
\nonumber &&\cr \nonumber && \cr \nonumber &&\cr
\end{eqnarray}
\begin{eqnarray}
\nonumber
\end{eqnarray}
\begin{center}
\begin{Large}
\textbf{Exact Solution for the Second Harmonic Generation in
XFELs}
\end{Large}
\begin{eqnarray}
\nonumber &&\cr \nonumber && \cr
\end{eqnarray}

\begin{large}
Gianluca Geloni, Evgeni Saldin, Evgeni Schneidmiller and Mikhail
Yurkov
\end{large}
\textsl{\\Deutsches Elektronen-Synchrotron DESY, Hamburg}
\begin{eqnarray}
\nonumber
\end{eqnarray}
\begin{eqnarray}
\nonumber
\end{eqnarray}
\begin{eqnarray}
\nonumber
\end{eqnarray}
ISSN 0418-9833
\begin{eqnarray}
\nonumber
\end{eqnarray}
\begin{large}
\textbf{NOTKESTRASSE 85 - 22607 HAMBURG}
\end{large}
\end{center}
%\end{widetext}
\clearpage
\newpage

\begin{frontmatter}



\title{Exact Solution for the Second Harmonic Generation in XFELs}


\author[DESY]{Gianluca Geloni}
\author[DESY]{Evgeni Saldin}
\author[DESY]{Evgeni Schneidmiller}
\author[DESY]{Mikhail Yurkov}

\address[DESY]{Deutsches Elektronen-Synchrotron (DESY), Hamburg,
Germany}

\begin{abstract}
The generation of harmonic radiation through a non-linear
mechanism driven by bunching at fundamental frequency is an
important option in the operation of high gain Free-Electron
Lasers (FELs). The use of harmonic generation at a large scale
facility may result in achieving shorter radiation wavelengths for
the same electron beam energy. This paper describes a theory of
second harmonic generation in planar undulators with particular
attention to X-Ray FELs (XFELs). Our study is based on an exact
analytical solution of Maxwell equations, derived with the help of
the Green's function method. On the contrary, up-to-date
theoretical understanding of the second harmonic generation is
only limited to some estimation of the total radiation power based
on the source part of the wave equation. Moreover, we find that
such part of the wave equation is presented with several incorrect
manipulations among which is the omission of an important
contribution. Our work yields correct parametric dependencies and
specific predictions of additional properties such as
polarization, angular distribution of the radiation intensity and
total power. The most surprising prediction is the presence of a
vertically polarized part of the second harmonic radiation,
whereas up-to-date understanding assumes that the field is
horizontally polarized. Altogether, this paper presents the first
correct theory of second harmonic generation for high gain FELs.
\end{abstract}


\end{frontmatter}

% main text

\clearpage
\section{\label{sec:intro} Introduction}

In a Free-Electron Laser the electromagnetic field at the
fundamental harmonic interacts with the electron beam. As a
result, the beam is bunched in non-linear (sinusoidal)
ponderomotive potential. When the bunching is strong enough, the
beam current exhibits non-negligible Fourier components at
harmonics of the fundamental as well. In the SASE case only the
transverse ground mode of the fundamental harmonic survives, due
to the transverse mode selection mechanism \cite{FELP} in the high
gain regime, and is responsible for the bunching mechanism. As a
result, the nonlinear Fourier components radiate coherently and
the phenomenon is referred to as (nonlinear) harmonic generation
of coherent radiation.

The process of harmonic generation of coherent radiation can be
considered as a purely electrodynamical one. In fact, the
harmonics of the electron beam density are driven by the
electromagnetic field at the fundamental frequency, but the
bunching contribution due to the interaction of the electron beam
with the radiation at higher harmonics can be neglected. This
leads to important simplifications. In fact, in order to perform
numerical analysis of the characteristics of higher harmonics
radiation, one has to solve the self-consistent problem for the
fundamental harmonic only. Subsequently, the solution to this
problem, that must be obtained with the help of a self-consistent
code, can be used to calculate the harmonic contents of the beam
current. These contents enter as known sources in our
electrodynamical process: solving Maxwell equations accounting for
these sources gives the desired characteristics of higher
harmonics radiation. As a result, simulation codes dealing with
harmonic generation are not first principle codes: in fact, they
simply compute the solution of Maxwell equations obtaining the
proper sources by means of first-principle codes.

Non-linear generation of the second harmonic radiation, in
particular, is important for extending the attainable frequency
range of an XFEL facility: in fact, the high peak-brilliance
increase of XFELs with respect to third generation light sources
(up to eight orders of magnitude) makes the second harmonic
contents of the XFEL radiation very attractive from a practical
viewpoint. Moreover, it is also important in connection with
experiments that make use of the fundamental harmonic only. In
fact one must be able to estimate correctly the higher harmonics
effects to distinguish between nonlinear phenomena induced by the
fundamental and linear phenomena due to the second harmonic.

The subject has been a matter of theoretical studies in a
high-gain SASE FEL both for odd \cite{KIM1} and even harmonics
\cite{SCHM,KIM2,RHUA}, where the electrodynamical problem is dealt
with. The practical interest of these studies is well underlined
by the fact that they were followed by both numerical analysis
\cite{FREU} and experiments, that have been carried out in the
infra-red and in the visible range of the electromagnetic spectrum
\cite{TREM,BIED}.

Experimental results are compared with numerical analysis and
numerical analysis rely on analytical studies: this fact stresses
the importance of a correct theoretical understanding of the
subject. Remarkably, such understanding does not require the
introduction of radically new physical mechanisms. The key ideas
involved are not much different from those regarding the second
harmonic generation from a single particle, treated a long time
ago and presented in Synchrotron Radiation textbooks (e.g.
\cite{WIE2,UNDU}). A complexity though, is constituted by the
presence of many electrons involved in the radiative process, each
with a given offset and phase, radiating coherently, as a whole or
in part, due to the longitudinal modulation of the beam current at
the second harmonic.

For a given frequency component the electromagnetic wave equation
dictates both a characteristic longitudinal length (that is the
radiation formation length) and a characteristic transverse
length. As we will see, when the beam transverse size is smaller
than the characteristic transverse length the entire electron beam
behaves like a single electron and the harmonics of the beam
current are simply interpretable in terms of the harmonic contents
of a single particle current: in this case, all the particles act
coherently and the radiated intensity scales with the square
number of the electrons in the beam. When the transverse size of
the beam increases and becomes much larger than the characteristic
transverse length less electrons contribute collectively to the
field and if the beam current remains constant the total radiated
power is decreased.

The characteristic transverse length is specified in a natural way
after a dimensional analysis of the problem. The first treatment
\cite{SCHM} of non-linear generation of even harmonics does not
account for the presence of such parameter, followed in this by
others \cite{KIM2, RHUA}.  We find that these works include
arbitrary manipulations of the source terms in the paraxial wave
equation. Among these, an important part of the source terms is
systematically dropped. Moreover, estimations of the second
harmonic power are based on the electromagnetic sources (after
manipulation) while exact calculations should be based on a
solution of Maxwell equations. Altogether, we find that these
works predict an incorrect dependence of the second harmonic field
on the problem parameters. Results of \cite{KIM2} are extended in
\cite{RHUA} to the case of an electron beam moving off-axis
through the undulator. One of the conclusions in \cite{RHUA} is
that the second harmonic power increases when an angle between the
beam and the undulator axis is present. We find that the power of
the second harmonic radiation should never increase when such
angle is present: in particular, as we will see, it is independent
of it in optimal situations when the microbunching wavefront is
matched with the beam propagation.

In this paper, that was inspired by a method \cite{OURS} developed
to deal with Synchrotron Radiation from complex setups, we present
a theory of second harmonic generation in high-gain FELs. First we
give, in Section \ref{sec:ours}, an exact analytical solution of
the wave equation for the second harmonic generation problem. The
procedure employed to derive such a solution shows the advantages
of a Green's function method. In Section \ref{sec:mod}, our result
is used to calculate, in a particular case, specific properties of
the second harmonic radiation such as polarization, directivity
diagram and total power including proper parametric dependencies.
The most surprising prediction of our theory is that the electric
field is not only horizontally polarized, as it is usually
assumed, but exhibits, though remaining linearly polarized, a
vertically polarized component too. Following the presentation of
our theory, in Section \ref{sec:stat} we comment on the
differences between our approach and the present understanding of
the second harmonic generation mechanism. Finally, in Section
\ref{sec:conc}, we come to conclusions.

\section{\label{sec:ours} Complete analysis of Second Harmonic Generation mechanism }

As has been said in the Introduction, the process of (second)
harmonic generation of coherent radiation is a purely
electrodynamical one. First, proper initial conditions are given
as input to an FEL self-consistent code, which calculates the
electron beam bunching from the interaction of the beam with the
first harmonic radiation. Then, the results from the
self-consistent code are used as electromagnetic sources to solve
the problem of second harmonic generation. For simplicity, in the
following we will consider a beam modulated at a single frequency
$\omega$ as the source. One may always write the longitudinal
current density $j_z$ along the undulator as a sum of an
unperturbed part independent of the modulation and of the time,
$j_{o}$, and a term responsible for the beam modulation,
$\tilde{j}_{z}$, at frequency $\omega$ (perturbation):

\begin{equation}
j_z(z,\vec{r}_\bot,t) = j_{o}(z,\vec{r}_\bot) +
\tilde{j}_{z}(z,\vec{r}_\bot,t)~. \label{sum1}
\end{equation}
We assume that we can write the unperturbed part $j_{o}$ as if all
the particles where moving coherently, that is

\begin{equation}
j_{o}(z,\vec{r}_\bot) = {j}_{o}(\vec{r}_\bot -
\vec{r}^{(c)}_\bot(z))~, \label{unp}
\end{equation}
where $\vec{r}^{(c)}_\bot(z)$ describes the coherent motion. This
assumption is always verified, for instance, in the case of a
single particle, when $j_o$ is simply a $\delta$-Dirac function,
or in the case of a monochromatic beam. If some energy spread is
present, in order for Eq. (\ref{unp}) to be valid we should assume
that the transverse size of the electron beam is not smaller than
the typical wiggling motion of the electrons. In this case, the
validity of Eq. (\ref{unp}) has an accuracy given by the relative
deviation of the particles energy form the average value, $\delta
\gamma/\gamma$. Since for the FEL process $\delta \gamma/\gamma$
is, at most, of the order of the efficiency parameter, we have
$\delta \gamma/\gamma \ll 1$ and Eq. (\ref{unp}) is valid with the
same accuracy of FEL theory. However it should be noted here that
the average energy of the beam is to be considered, in general, a
function of the coordinate $z$, $\gamma=\gamma(z)$: it has to be
given as a result of start-to-end simulations and considered as an
input for our electromagnetic problem.

%\footnote{We will not discuss, here,
%exotic schemes where, for instance, an acceleration module is
%installed between two undulator sections.}. Therefore, for the
%purpose of calculating $\vec{r'}_\bot(z)$, the average energy of
%the beam, which is, in general, a given function $\gamma =
%\gamma(z)$, can be considered as a known constant $\gamma \simeq
%\bar{\gamma} = \mathrm{const}$.

The perturbation $\tilde{j}_{z}$ can then be written as

\begin{eqnarray}
\tilde{j}_{z}(z,t) &=&
j_{o}\left(\vec{r}_\bot-\vec{r}^{(c)}_\bot(z)\right) \cr &&\times
\left\{\tilde{a}_2\left(z,\vec{r}_\bot-\vec{r}^{(c)}_\bot(z)\right)\exp\left[i
\omega \int_0^z \frac{dz'}{v_z(z')} - i \omega t
\right]+\mathrm{C.C.}\right\}~. \label{jzp}
\end{eqnarray}
The function $\tilde{a}_2$ is to be considered a result from the
FEL self-consistent code, and its dependence on $z$ describes the
evolution of the modulation through the beamline and accounts for
emittance and energy spread effects. It should be noted that the
values of $\tilde{a}_2$ are not necessarily real: in fact there
can be a $z$-dependent phase shift with respect to the phase
$\omega \int_0^z {dz'}/v_z(z')-\omega t$.

In order to correctly calculate the phase $\omega \int_0^z
{dz'}/v_z(z')-\omega t$ in Eq. (\ref{jzp}) one has to account for
the dependence of the longitudinal velocity associated with the
coherent motion, $v_z$ on the position $z$. The function
${v}_z(z)$ can be recovered from the knowledge of
$\vec{r}^{(c)}_\bot(z)$ and of the average energy of the beam
$\gamma=\gamma(z)$.

%In this case, contrary to what has been said before regarding the
%calculation of $\vec{r'}_\bot(z)$, $\gamma=\gamma(z)$ cannot be
%approximated with $\bar{\gamma}$: it has to be given as a result
%of start-to-end simulations and considered as an input for our
%electromagnetic problem.

If the beam is deflected of angles $\eta_x$ and $\eta_y$ in the
horizontal and vertical direction with respect to the $z$ axis,
the velocity of the coherent motion depends also on the deflection
angles. Renaming position and velocity of the coherent motion with
no deflection with the subscript "(nd)" one obtains:

\begin{eqnarray}
v_z(z,\eta) &=& v_{z(nd)}(z)
\left(1-\frac{\eta_x^2+\eta_y^2}{2}\right) \cr
\vec{v}_\bot(z,\eta) &=& \vec{v}_{\bot (nd)}(z) + v_{z(nd)}(z)
\vec{\eta}~, ~~~~\label{etav}
\end{eqnarray}
and

\begin{eqnarray}
\vec{r}^{(c)}_\bot(z,\vec{\eta}) = \vec{r}^{(c)}_{\bot(nd)}(z) +
\vec{\eta} z ~. ~~~~\label{etar}
\end{eqnarray}
Also $\tilde{a}_2$ will depend on $\vec{\eta}$. The exact
dependence is fixed by the way the beam is prepared and should be
regarded as a condition for the orientation of the microbunching
wavefront. In all generality we can write:

\begin{equation}
\tilde{a}_2 =
\tilde{a}_2\left(z,\vec{r}_\bot-\vec{r}^{(c)}_\bot(z,\vec{\eta})\right)
~.\label{roto}
\end{equation}
In the limit for $\gamma^2 \gg 1$, the total current density can
be written as

\begin{eqnarray}
\vec{j}(z,t,\vec{\eta})&=&\frac{\vec{v}(z,\vec{\eta})}{c}
j_{o}\left(\vec{r}_\bot-\vec{r}^{(c)}_\bot(z,\vec{\eta})\right)
\times \Bigg\{1+\Bigg[\tilde{a}_2\left(z,
\vec{r}_\bot-\vec{r}^{(c)}_\bot(z,\vec{\eta})\right)\cr&&\exp\Bigg[i
\omega \int_0^z \frac{dz'}{v_z(z',\vec{\eta})}-i\omega t
\Bigg]+\mathrm{C.C.}\Bigg]\Bigg\}~. \label{totcur}
\end{eqnarray}
%
%Finally, due to fast oscillations in time one has
%
%\begin{eqnarray}
%\overline{\int j_z d \vec{r}_\bot}= \int j_o d \vec{r}_\bot =
%\mathrm{const},  \label{conse}
%\end{eqnarray}
%%
%where $\overline{(...)}$ indicates time averaging over one period
%o the fast oscillations in time and the integrals are to be
%performed over a vertical plane at any given position in $z$. As a
%result, because of charge conservation,

One can express the charge density as

\begin{equation}
\rho =\frac{j_z}{v_z} \simeq \frac{j_z}{c}~, \label{chd}
\end{equation}
as we will be working in the paraxial approximation.

Eq. (\ref{totcur}) and Eq. (\ref{chd}) give us the expressions to
be used as sources for Maxwell equation. Looking for solutions for
$\vec{E}_\bot$ in the form

\begin{equation}
\vec{{E}}_\bot = \vec{\widetilde{E}}_\bot \exp\left[i\omega
(z/c-t)\right] + \mathrm{C.C.}~\label{etildadef}
\end{equation}
and applying the paraxial approximation, one may write the Maxwell
equation describing $\vec{\widetilde{E}}_\bot$ as \cite{OURS}:

\begin{eqnarray}
\left({\nabla_\bot}^2 + \frac{2 i \omega}{c}
\frac{\partial}{\partial z}\right) \vec{\widetilde{E}}_{\bot} =
\frac{4 \pi}{c}  \exp\left[i \left(\Phi_s-\omega
\frac{z}{c}\right)\right] \left[\frac{i\omega}{c^2}\vec{v}_\bot
-\vec{\nabla}_\bot \right] j_o \tilde{a}_2 ~, \label{incipit4}
\end{eqnarray}
were we have put

\begin{equation}
\Phi_s(z,\vec{\eta}) = \omega \int_0^{z} \frac{d
z'}{v_z(z',\vec{\eta})} ~. \label{phu}
\end{equation}
With the aid of the appropriate Green's function an exact solution
of Eq. (\ref{incipit4}) can be found without any extra assumption
about the parameters of the problem.

\begin{eqnarray}
\widetilde{\vec{E}}_{\bot }(z_o, \vec{r}_{\bot o} )&=&
-\frac{1}{c}\int_{-\infty}^{\infty} dz' \frac{1}{z_o-z'} \int d
\vec{r'}_{\bot}
\left[\frac{i\omega}{c^2}\vec{v}_\bot(z',\vec{\eta})
-\vec{\nabla}'_\bot \right]\cr &&\times
j_{o}\left(\vec{r'}_\bot-\vec{r}^{(c)}_\bot(z',\vec{\eta})\right)
\tilde{a}_2 \left(z',\vec{r'}_\bot
-\vec{r}^{(c)}_\bot(z',\vec{\eta}) \right)\cr&&
\exp\left\{i\omega\left[\frac{\mid \vec{r}_{\bot o}-\vec{r'}_\bot
\mid^2}{2c (z_o-z')}\right]+ i \left[ \Phi_s(z',\vec{\eta})-\omega
\frac{z'}{c}\right] \right\} ~, \label{blob}
\end{eqnarray}
where $\vec{\nabla}'_\bot$ represents the gradient operator with
respect to the source point, while $(z_o, \vec{r}_{\bot o})$
indicates the observation point. Integration by parts of the
gradient terms leads to

\begin{eqnarray}
\widetilde{\vec{E}}_{\bot}&= &-\frac{i \omega }{c^2}
\int_{-\infty}^{\infty} dz' \frac{1}{z_o-z'}  \int d
\vec{r'}_{\bot} \left(\frac{\vec{v}_\bot(z',\vec{\eta})}{c}
-\frac{\vec{r}_{\bot o}-\vec{r'}_\bot}{z_o-z'}\right)\cr&&\times
j_{o}\left(\vec{r'}_\bot-\vec{r}^{(c)}_\bot(z',\vec{\eta})\right)
\tilde{a}_2 \left(z',\vec{r'}_\bot
-\vec{r}^{(c)}_\bot(z',\vec{\eta}) \right) \exp\left[i
\Phi_T(z',\vec{r'}_\bot,\vec{\eta})\right] ~, \cr &&
\label{generalfin}
\end{eqnarray}
where the total phase $\Phi_T$ is given by

\begin{equation}
\Phi_T =  \left[\Phi_s-\omega\frac{z'}{c}\right]+ \omega \left[
\frac{|\vec{r}_{\bot o}-\vec{r'}_\bot|^2}{2c (z_o-z')}\right]~.
\label{totph}
\end{equation}
We will now make use of a new integration variable $\vec{l}=
\vec{r'}_\bot-\vec{r}^{(c)}_\bot(z',\vec{\eta})$ so that

\begin{eqnarray}
\widetilde{\vec{E}}_{\bot }&=& -\frac{i \omega }{c^2}
\int_{-\infty}^{\infty} dz' \frac{1}{z_o-z'}  \int d \vec{l}
\left(\frac{\vec{v}_\bot(z',\vec{\eta})}{c} -\frac{\vec{r}_{\bot
o}- \vec{r}^{(c)}_\bot(z',\vec{\eta})-\vec{l}}{z_o-z'}\right)\cr&&
\times j_{o}\left(\vec{l}\right) \tilde{a}_2 \left(z',\vec{l}
\right) \exp\left[i \Phi_T(z',\vec{l},\vec{\eta})\right] ~,
\label{generalfin2}
\end{eqnarray}
and

\begin{equation}
\Phi_T =  \left[\Phi_s-\omega\frac{z'}{c}\right]+ \omega \left[
\frac{|\vec{r}_{\bot
o}-\vec{r}^{(c)}_\bot(z',\vec{\eta})-\vec{l}|^2}{2c
(z_o-z')}\right]~. \label{totph2}
\end{equation}
We will consider the case of a planar undulator and we will be
interested in the total power of the second harmonic emission and
in the directivity diagram of the radiation in the far zone.
Accounting for the beam deflection angles $\eta_x$ and $\eta_y$ we
model the electron transverse motion as:

\begin{equation}
\vec{v}_\bot(z',\vec{\eta}) = \left[- {c K\over{\gamma}}
\sin{\left(k_w z'\right)}+\eta_x v_z\right] \vec{x}+\left[\eta_y
v_z \right]\vec{y}~, \label{vuz2}
\end{equation}
and

\begin{eqnarray}
\vec{r}^{(c)}_\bot(z',\vec{\eta})+\vec{l} = \left[ \frac{K}{\gamma
k_w} \left(\cos{\left(k_w z'\right)}-1\right)+\eta_x z'
+l_x\right] \vec{x}  + \left[\eta_y z' +l_y\right]\vec{y}~.
\label{erz2}
\end{eqnarray}
Here $K$ is the deflection parameter and $k_w = 2\pi/\lambda_w$,
$\lambda_w$ being the undulator period. Moreover, one has

\begin{eqnarray}
\frac{c \Phi_s}{\omega}  &\simeq& \left(\frac{4
\gamma^2}{4\gamma^2 - K^2 }+\frac{\eta_x^2+\eta_y^2}{2}\right)z' -
\frac{K \eta_x}{k_w \gamma} \cr&& - \frac{K^2}{8\gamma^2
k_w}\sin\left(2 k_w z' \right)+ \frac{K \eta_x}{\gamma
k_w}\cos\left(k_w z' \right)~, \label{zsss2}
\end{eqnarray}
We will now introduce the far zone approximation. Substitution of
Eq. (\ref{zsss2}), Eq. (\ref{erz2}) and Eq. (\ref{vuz2}) in Eq.
(\ref{generalfin}) yields the field contribution calculated along
the undulator:

\begin{eqnarray}
\widetilde{\vec{E}}_{\bot} &=& \frac{i \omega }{c^2 z_o} \int
d\vec{l} \int_{-L_w/2}^{L_w/2} dz' j_{o}\left(\vec{l}\right)
\tilde{a}_2 \left(z',\vec{l} \right){\exp{\left[i \Phi_T\right]}}
\cr &&\times \left[\left(\frac{K}{\gamma} \sin\left(k_w
z'\right)+\left(\theta_x-\eta_x\right)\right)\vec{{x}}
+\left(\theta_y-\eta_y\right)\vec{{y}}\right]~, \label{undurad}
\end{eqnarray}
where

\begin{eqnarray}
\Phi_T &=& {\omega} \left\{ \frac{z'}{2\gamma^2 c}
\left[1+\frac{K^2}{2} +
\gamma^2\left((\theta_x-\eta_x)^2+(\theta_y-\eta_y)^2\right)\right]\right.
\cr&&\left.-\frac{K^2}{8\gamma^2k_w c} \sin{(2 k_w z')} - \frac{K
(\theta_x-\eta_x)}{\gamma k_w c} \cos{(k_w z')}\right\} \cr&&
+\omega \left\{\frac{K}{k_w\gamma c}(\theta_x-\eta_x)
-\frac{1}{c}(\theta_x l_x+ \theta_y l_y) + (\theta_x^2+\theta_y^2)
\frac{z_o}{2c} \right\} ~.\label{phitundu}
\end{eqnarray}
Here $\theta_{x}$ and $\theta_{y}$ indicate the observation angles
$x_o/z_o$ and $y_o/z_o$. Moreover, the integration is performed in
from $-L_w/2$ to $L_w/2$ in $d {z}'$, $L_w = N_w \lambda_w$ being
the undulator length. In fact, working under the resonance
approximation in the limit for $N_w \gg 1$ allows us to neglect
contributions outside the undulator \cite{OURS}.

We will make use of the well-known expansion (see \cite{ALFE})

\begin{equation}
\exp{[i a \sin{(\psi)}]}=\sum_{p=-\infty}^{\infty} J_p(a) \exp{[i
p \psi]}~, \label{alfeq} \end{equation}
where $J_p$ indicates the Bessel function of the first kind of
order $n$.

We will be interested in frequencies around the second harmonic:

\begin{equation}
\omega_{2o} = 4 {k_w c {\gamma}_z^2} ~,\label{freqfix}
\end{equation}
where

\begin{equation}
\gamma_z^2 = \frac{\gamma^2}{1+K^2/2}~. \label{gammaz}
\end{equation}
Indicating with $\widetilde{\vec{E}}_{\bot 2}$ the second harmonic
contribution calculated at frequencies around $\omega_{2o}$ one
obtains

\begin{eqnarray}
\widetilde{\vec{E}}_{\bot 2}&=& \frac{i \omega_{2o}  }{c^2 z_o}
\int_{-\infty}^{\infty} d l_x \int_{-\infty}^{\infty} d l_y
\int_{-L_w/2}^{L_w/2} dz' j_{o}\left(\vec{l}\right) \tilde{a}_2
\left(z',\vec{l} \right) \exp[i\Phi_o] \cr &&\times
\sum_{m=-\infty}^{\infty} \sum_{n=-\infty}^{\infty} J_m(u)
J_{n}(v)  \exp\left[\frac{i n \pi}{2}\right] \cr && \times
\Bigg\{\Bigg[-i \frac{K}{2\gamma}
\Bigg(\exp\left\{i[R_\omega+1]k_w z'\right\}
-\exp\left\{i[R_\omega-1]k_w z'\right\}\Bigg)
\cr&&+(\theta_x-\eta_x) \exp\left\{i R_\omega k_w z'\right\}
\Bigg]\vec{x} +\Bigg[(\theta_y-\eta_y) \exp\left\{i R_\omega k_w
z'\right\} \Bigg]\vec{y}\Bigg\} ~, \label{undurad2}
\end{eqnarray}
where

\begin{equation}
R_\omega = \frac{\omega}{\omega_1} - n -2 m ~,\label{romega}
\end{equation}
with

\begin{equation}
\omega_1^{-1}=
\frac{1}{2k_wc\gamma^2}\left\{1+\frac{K^2}{2}+\gamma^2\left[
\left(\theta_x-\eta_x\right)^2+
\left(\theta_y-\eta_y\right)^2\right]\right\}\label{omega1}~.
\end{equation}
Moreover

\begin{equation}
u=\frac{\omega_{2o}}{\omega_1} \frac{K^2
\left[1-K^2/(4\gamma^2)\right]}
{4\left\{1+\frac{K^2}{2}+\gamma^2\left[
\left(\theta_x-\eta_x\right)^2+
\left(\theta_y-\eta_y\right)^2\right]\right\}}\label{v},~
\end{equation}
\begin{equation}
v=\frac{\omega_{2o}}{\omega_1} \frac{2K  \gamma
\left[1-K^2/(4\gamma^2)\right](\theta_x-\eta_x)}
{1+\frac{K^2}{2}+\gamma^2\left[ \left(\theta_x-\eta_x\right)^2+
\left(\theta_y-\eta_y\right)^2\right]} \label{u}~
\end{equation}
and
\begin{equation}
\Phi_o = \omega_{2o} \left[\frac{K}{k_w\gamma c} (\theta_x-\eta_x)
- \frac{1}{c}(\theta_x l_x+\theta_y
l_y)+\frac{z_o}{2c}(\theta_x^2+\theta_y^2) \right]~. \label{phio}
\end{equation}
In the limit for $N_w \gg 1$ and if $\tilde{a}_2$ does not vary
much in $z'$ over a period of the undulator $\lambda_w$ the fast
oscillations in the exponential function in the integrand of Eq.
(\ref{undurad2}) tend to suppress the integral unless $R_\omega =
0$, $R_\omega=-1$  or $R_\omega = 1$, that is when at least one of
the exponential function is simply unity. If $\omega = \omega_2$
this corresponds to  $n = 2 - 2m$, $n=3-2 m $ and $n=1-2 m $
respectively.  Neglecting all other terms and imposing $\omega =
\omega_2 + \Delta \omega_2$ we obtain

\begin{eqnarray}
\widetilde{\vec{E}}_{\bot 2} &=& \frac{i \omega_{2o}  }{c^2 z_o}
\int_{-\infty}^{\infty} d l_x \int_{-\infty}^{\infty} d l_y
\int_{-L_w/2}^{L_w/2} dz' j_{o}\left(\vec{l}\right) \tilde{a}_2
\left(z',\vec{l} \right) \exp[i\Phi_o] \exp\left\{i\frac{\Delta
\omega_2}{\omega_1} k_w z'\right\}  \cr &&
\times\sum_{m=-\infty}^{\infty} \Bigg\{\Bigg[-i \frac{K}{2\gamma}
\left( J_m(u) J_{3-2m}(v) \exp\left\{\frac{i [3-2m]
\pi}{2}\right\} \right.\cr&&\left. -J_m(u)  J_{1-2m}(v)
\exp\left\{\frac{i [1-2m] \pi}{2}\right\}\right) \cr&&
+(\theta_x-\eta_x) J_m(u) J_{2-2m}(v) \exp\left\{\frac{i [2-2m]
\pi}{2}\right\} \Bigg]\vec{x} \cr&&+\Bigg[(\theta_y-\eta_y) J_m(u)
J_{2-2m}(v) \exp\left\{\frac{i [2-2m]\pi}{2}\right\}
\Bigg]\vec{y}\Bigg\} ~. \label{undurad3}
\end{eqnarray}
For any value of $K$ and $\theta_{x}-\eta_x$ much smaller than
$1/\gamma_z$, $v$ is a small parameter and only the smallest
indexes in the Bessel functions $J_q(v)\sim v^{q}$ in Eq.
(\ref{undurad2}) give non negligible contribution. As a result,
Eq. (\ref{undurad2}) can be drastically simplified. One can write
(compare, for instance, with \cite{OURS}):

\begin{eqnarray}
\widetilde{\vec{E}}_{\bot 2}&=&\frac{i \omega_{2o} }{c^2 z_o}
\left[\mathcal{A}(\theta_x-\eta_x)\vec{x} +
\mathcal{B}(\theta_y-\eta_y)\vec{y}\right] \int_{-\infty}^{\infty}
d l_x \int_{-\infty}^{\infty} d l_y \int_{-L_w/2}^{L_w/2} d z'
\exp{[i\Phi_o]} \cr && \times j_{o}\left(\vec{l}\right)
\tilde{a}_2 \left(z',\vec{l} \right) \exp[i C z'] \exp\left\{i 2
\gamma_z^2 \left[\left(\theta_x-\eta_x\right)^2+
\left(\theta_y-\eta_y\right)^2\right] k_w z'\right\}
\cr&&\label{undurad4}
\end{eqnarray}
where we have defined

\begin{equation}
\mathcal{A} = \frac{2
K^2}{2+K^2}\left[J_0\left(\frac{K^2}{2+K^2}\right)-J_2
\left(\frac{K^2}{2+K^2}\right)\right] +
J_1\left(\frac{K^2}{2+K^2}\right)~, \label{calA}
\end{equation}
\begin{equation}
\mathcal{B} = J_1\left(\frac{K^2}{2+K^2}\right)~, \label{calB}
\end{equation}
we have used the fact that

\begin{equation}
\frac{ \Delta \omega_2}{\omega_1} = 2 \gamma_z^2
\left[\left(\theta_x-\eta_x\right)^2+
\left(\theta_y-\eta_y\right)^2\right]+C~,\label{argument}
\end{equation}
with

\begin{equation}
C = \frac{\omega - \omega_{2o}}{\omega_{1o}} \label{Cdef}
\end{equation}
and, under the resonant approximation, we have

\begin{equation}
\Phi_o = \frac{\omega_{2o}}{c} \left[ - (\theta_x l_x+\theta_y
l_y)+\frac{z_o}{2}(\theta_x^2+\theta_y^2) \right]~.
\label{phiosimpli}
\end{equation}
\begin{figure}
\begin{center}
\includegraphics*[width=140mm]{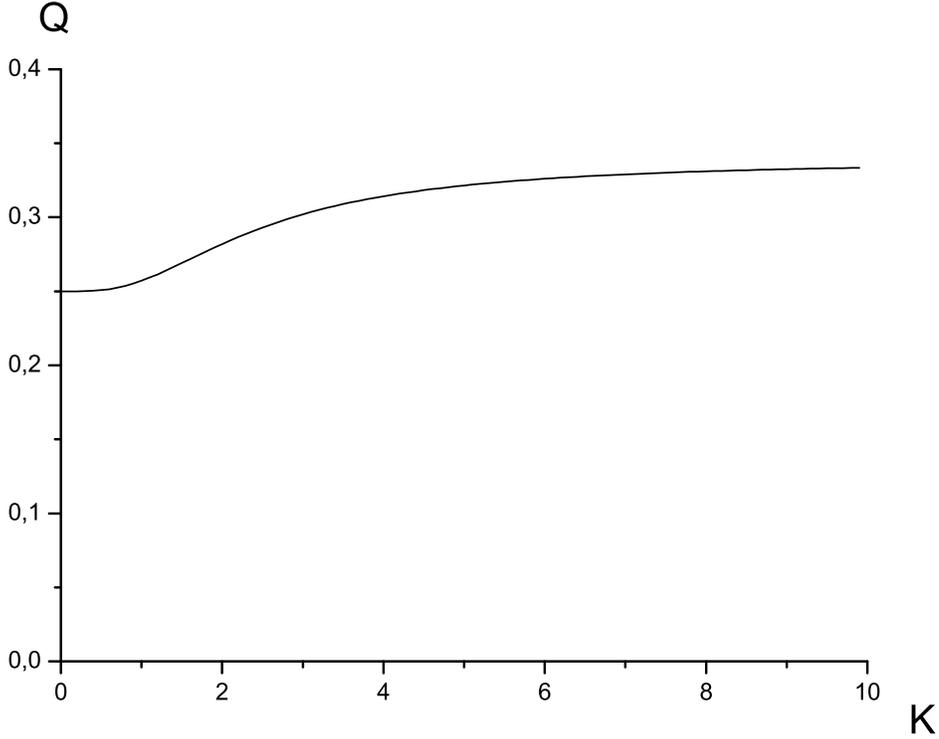}% Here is how to import EPS art
\caption{\label{Ratio} Illustration of the behavior of the ratio
$Q(K)$ between the second harmonic field contribution due to the
gradient of the density part of the source and the contribution
due to the current part of the source.}
\end{center}
\end{figure}
The detuning parameter $C$ should indeed be considered as a
function of $z$, $C=C(z)$ which can be retrieved from the
knowledge of $\gamma=\gamma(z)$.

It is important to see that the terms in $J_1$ in Eq. (\ref{calA})
and Eq. (\ref{calB}) are due to the presence of the gradient term
in $\vec{\nabla}_\bot (j_o \tilde{a}_2)$ in Eq. (\ref{incipit4}),
which has been omitted in \cite{SCHM} and later on in  \cite{KIM2,
RHUA}. We find that, without the gradient, term one would recover
results quantitatively incorrect for the $x$-polarization
component. In Fig. \ref{Ratio} we plotted the ratio between the
contribution of the radiation field due to the gradient of the
density part of the source and the contribution due to the current
part of the source for the $x$-polarization component. This is a
function $Q(K)$ of the $K$ parameter only and it can be written as

\begin{eqnarray}
Q = \frac{\tilde{E}_{\bot 2 g}}{\tilde{E}_\mathrm{\bot 2 c}}  =
\frac{2+K^2}{2 K^2} J_1\left(\frac{K^2}{2+K^2}\right) \Bigg/
\left[J_0\left(\frac{K^2}{2+K^2}\right)-J_2
\left(\frac{K^2}{2+K^2}\right)\right]&&~,\cr && \label{Ratioeq}
\end{eqnarray}
where the subscript "g" stands for "gradient and "c" stands for
"current". As it can be seen from Fig. \ref{Ratio}, the gradient
term always contributes for more than one fourth of the total
field, independently of the values of $K$. Also, if the gradient
term is omitted, the entire contribution to the field polarized in
the $y$ direction would go overlooked. The inclusion of the
gradient term in the source part of the wave equation should not
be considered as a peculiarity of the second harmonic generation
mechanism. In Synchrotron Radiation theory from bending magnets,
for instance, the presence of such a source term is customary and
it is responsible, as here, for part of the horizontally polarized
field and for the entire vertically polarized field. Moreover, the
gradient term is always associated with an integration by part,
and therefore is always accompanied with the gradient of the
Green's function, which is responsible for a term proportional to
the observation angle $\theta_{x,y}$.

Eq. (\ref{undurad4}) can be also written as:

\begin{eqnarray}
\tilde{\vec{E}}_{\bot 2}&=&\frac{i \omega_{2o}} {c^2 z_o}
\exp{\left[i \frac{\omega_{2o}}{2c}
{z}_o({\theta}_x^2+{\theta}_y^2) \right]}
\left[\mathcal{A}({\theta}_x-{\eta}_x)\vec{x} +
\mathcal{B}({\theta}_y-{\eta}_y)\vec{y}\right]\cr&&
\times\int_{-\infty}^{\infty} d {l}_x \int_{-\infty}^{\infty} d
{l}_y \int_{-\infty}^{\infty} d {z}' \exp{\left[-i
\frac{\omega_{2o}}{c} \left({\theta}_x l_x + \theta_y
l_y\right)\right] } \cr &&\times \exp\left\{i \frac{\omega_{2o}}{2
c} \left[\left(\theta_x-\eta_x\right)^2+
\left(\theta_y-\eta_y\right)^2\right] z'\right\}
\tilde{\rho}^{(2)}({z}',\vec{{l}},{C}) ~,\label{undurad5}
\end{eqnarray}
where we have defined $\tilde{\rho}$ as

\begin{equation}
\tilde{\rho}^{(2)}({z}',\vec{l},{C}) = j_{o}\left(\vec{l}\right)
\tilde{a}_2 \left(z',\vec{l} \right) \exp\left[i C
z'\right]H_{L_w}(z') ~,\label{rhodefin}
\end{equation}
$H_{L_w}(z')$ being a function equal to unity over the interval
$[-L_w/2,L_w/2]$ and zero everywhere else. Its introduction simply
amounts to a notational change. Namely it accounts for the fact
that the integral in $dz'$ is performed over the undulator length
in Eq. (\ref{undurad4}), while it is performed from $-\infty$ to
$\infty$ in Eq. (\ref{undurad5}).  It should be noted that,
usually, computer codes do not present the functions $\tilde{a}_2$
and $\exp[i C z']$ separately as we did, but rather they combine
them in a single product, usually known as the complex amplitude
of the electron beam modulation with respect to the phase $\psi =
2 k_w z' + (\omega/c) z' - \omega t$. Regarding
$\tilde{\rho}^{(2)}$ as a given function allows one not to bother
about a particular presentation of the beam modulation.

Eq. (\ref{undurad4}) or, equivalently, Eq. (\ref{undurad5}) are
our most general result, and are valid independently on the model
chosen for the current density and the modulation. It is
interesting to note here that, when one writes Eq.
(\ref{undurad4}) in the form of Eq. (\ref{undurad5}), one obtains
an expression which is \textit{formally} similar to the spatial
Fourier transform of $\tilde{\rho}^{(2)}({z}',\vec{{l}},{C})$ with
respect to ${z}'$ and $\vec{{l}}$. There are two problems though:
first, $\tilde{\rho}^{(2)}$ is a function of $\vec{\eta}$, which
appears in the conjugate variable to $z'$ and, second, if $\gamma
= \gamma(z)$ one has $\omega_{2o}=\omega_{2o}(z)$.

\section{\label{sec:mod} Analysis of a simple model}

Let us treat a particular case. Namely, let us consider the case
when we can consider $\gamma(z) = \bar{\gamma}= \mathrm{const}$,
when $C(z)=0$ and

\begin{equation}
\tilde{\rho}^{(2)}({z},\vec{l}) = j_{o}\left(\vec{l}\right) a_{2o}
\exp \left[i \frac{\omega_{2o}}{c} \left(\eta_x l_x + \eta_y l_y
\right) \right]H_{L_w}(z) ~, \label{expara}
\end{equation}
with $a_{2o} = \mathrm{const}$  and

\begin{equation}
{j}_{o}\left(\vec{{l}}\right) = \frac{I_o}{2\pi {\sigma^2}}
\exp{\left(-\frac{{l}_x^2+{l}_y^2}{2 \sigma^2}\right)}~,
\label{exbot}
\end{equation}
$I_o$ and $\sigma$ being the bunch current and transverse size
respectively.

This particular case corresponds to a modulation wavefront
perpendicular to the beam direction of motion. In this case Eq.
(\ref{undurad5}) can be written as

\begin{eqnarray}
\tilde{\vec{E}}_{\bot 2}&=&\frac{i a_{2o}\omega_{2o}} {c^2 z_o}
\exp{\left[i \frac{\omega_{2o}}{2c}
{z}_o({\theta}_x^2+{\theta}_y^2) \right]}
\left[\mathcal{A}({\theta}_x-{\eta}_x)\vec{x} +
\mathcal{B}({\theta}_y-{\eta}_y)\vec{y}\right]\cr&&
\times\int_{-\infty}^{\infty} d {l}_x \int_{-\infty}^{\infty} d
{l}_y \int_{-\infty}^{\infty} d {z}' \exp{\left\{-i
\frac{\omega_{2o}}{c} \left[\left({\theta}_x-\eta_x\right) l_x +
\left(\theta_y-\eta_y\right) l_y\right]\right\} } \cr &&\times
\exp\left\{i \frac{\omega_{2o}}{2 c}
\left[\left(\theta_x-\eta_x\right)^2+
\left(\theta_y-\eta_y\right)^2\right] z'\right\}
j_{o}\left(\vec{l}\right)  H_{L_w}(z')
 ~\label{undurad6}
\end{eqnarray}
and amounts, indeed to the spatial Fourier transform of
$j_o\left(\vec{l}\right)  H_{L_w}(z')$. We obtain
straightforwardly:

\begin{eqnarray}
\tilde{\vec{E}}_{\bot 2}&=& \frac{i I_o a_{2o}\omega_{2o} L_w}
{c^2 z_o} \exp{\left[i
\frac{\omega_{2o}}{2c}{z}_o({\theta}_x^2+{\theta}_y^2) \right]}
\left[\mathcal{A}({\theta}_x-{\eta}_x)\vec{x} +
\mathcal{B}({\theta}_y-{\eta}_y)\vec{y}\right] \cr&&
\times\mathrm{sinc} \left\{\frac{L_w \omega_{2o}}{4 c
}\left[\left( \theta_x - \eta_x\right)^2 + \left( \theta_y -
\eta_y\right)^2 \right]\right\}\cr &&\times
\exp{\left\{-\frac{\sigma^2 \omega_{2o}^2 }{2 c^2}\left[\left(
\theta_x - \eta_x\right)^2 + \left( \theta_y - \eta_y\right)^2
\right]\right\}} ~. ~\label{undurad8bis}
\end{eqnarray}
If the beam is prepared in a different way so that, for instance,
the modulation wavefront is not orthogonal to the direction of
propagation of the beam,  Eq. (\ref{undurad5}) retains its
validity. However it should be noted that, in this case, Eq.
(\ref{undurad5}) is not, in general, a Fourier transform. It is if
$\gamma(z) = \bar{\gamma}= \mathrm{const}$ and
$\tilde{\rho}^{(2)}$ includes a phase factor of the form
$\exp{\left[i \alpha \vec{\eta} \cdot \vec{l}\right]}$. The case
$\alpha = 1$ has just been treated. The case $\alpha = 0$
corresponds, instead, to a modulation wavefront orthogonal to the
$z$ axis and \textit{not} to the direction of propagation.

Going back to our particular case in Eq. (\ref{undurad8bis}), a
subject of particular interest is the angular distribution of the
radiation intensity along the $\vec{x}$ and $\vec{y}$ polarization
directions which will be denoted with $I_{2(x,y)}$. Upon
introduction of normalized quantities:

\begin{eqnarray}
\hat{\theta} &=& \sqrt{\frac{\omega_{2o} L_w}{c}}\theta=\sqrt{8\pi
N_w} \gamma_z \theta \cr \hat{\eta} &=&\sqrt{\frac{\omega_{2o}
L_w}{c}} \eta= \sqrt{8\pi N_w} \gamma_z \eta
 \cr
\hat{l}_{x,y}& =&\sqrt{\frac{\omega_{2o}}{c L_w}}l_{x,y}=
{\sqrt{8\pi N_w}} \frac{\gamma_z}{L_w} l_{x,y}  \label{reduced}
\end{eqnarray}
and of the Fresnel number:

\begin{equation}
N= \frac{\omega_{2o} \sigma^2}{c L_w} ~,\label{frsnintr}
\end{equation}
one obtains

\begin{eqnarray}
{I}_{2
(x,y)}\left(\hat{\theta}_{x}-\hat{\eta}_x,\hat{\theta}_{y}-\hat{\eta}_y
\right) &=& \mathrm{const} \times
\left(\hat{\eta}_{x,y}-\hat{\theta}_{x,y}\right)^2  \cr && \times
\mathrm{sinc}^2 {\left\{\frac{1}{4}\left[\left( \hat{\theta}_x -
\hat{\eta}_x\right)^2 + \left( \hat{\theta}_y -
\hat{\eta}_y\right)^2\right]\right\}} \cr&& \times
\exp{\left\{-N\left[\left( \hat{\theta}_x - \hat{\eta}_x\right)^2
+ \left( \hat{\theta}_y - \hat{\eta}_y\right)^2\right]\right\}}
~.\label{I2xybis}
\end{eqnarray}
Note that in the limit for $N \ll 1$, Eq. (\ref{I2xybis})
restitutes the directivity diagram for the second harmonic
radiation from a single particle. In agreement with Synchrotron
Radiation textbooks \cite{WIE2, UNDU} none of the polarization
components of ${I}_{2 (x,y)}$ has azimuthal symmetry, contrarily
with what happens for the first harmonic, where only the $x$
polarization is present and is endowed with azimuthal symmetry.

\begin{figure}
\begin{center}
\includegraphics*[width=140mm]{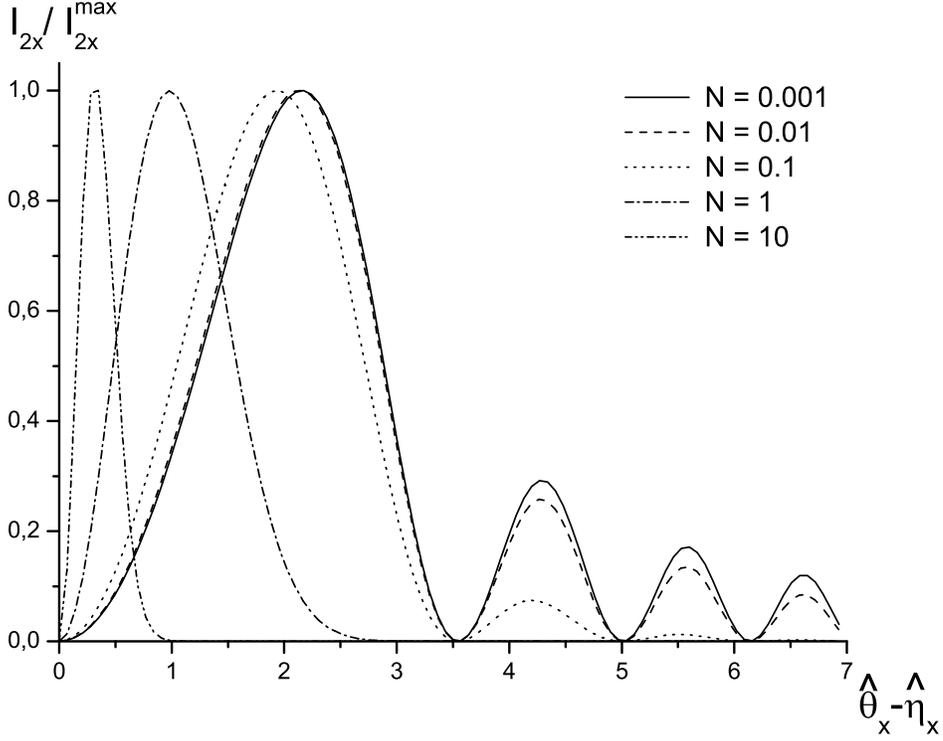}% Here is how to import EPS art
\caption{\label{DIRDIA} Plot of the directivity diagram for the
radiation intensity as a function of $\hat{\theta}_x -
\hat{\eta}_x$ at $\hat{\theta}_y-\hat{\eta}_y = 0$ for the
horizontal polarization component, for different values of ${N}$.}
\end{center}
\end{figure}
As an example, the directivity diagram in Eq. (\ref{I2xybis}) is
plotted in Fig. \ref{DIRDIA} for different values of ${N}$ as a
function of $\hat{\theta}_x - \hat{\eta}_x$ at
$\hat{\theta}_y-\hat{\eta}_y = 0$ for the horizontal polarization
component.

The next step is the calculation of the second harmonic power. The
power for the \textit{x-} and \textit{y-}polarization components
of the second harmonic radiation are given  by

\begin{eqnarray}
W_{2 (x,y)} = \frac{c}{4 \pi} \int_{-\infty}^{\infty} dx_o
\int_{-\infty}^{\infty} dy_o \overline{|E_{\bot x,y}(z_o, x_o,
y_o,t)|^2} &&\cr= \frac{c}{2 \pi} \int_{-\infty}^{\infty} dx_o
\int_{-\infty}^{\infty} dy_o {|\tilde{E}_{\bot x,y}(z_o, x_o,
y_o)|^2}~, \label{xpowden}
\end{eqnarray}
where $\overline{(...)}$ denotes averaging over a cycle of
oscillation of the carrier wave.

We will still consider the model specified by Eq. (\ref{expara})
and Eq. (\ref{exbot}) with $C=0$. It is convenient to present the
expressions for $W_{2x}$ and $W_{2y}$ in a dimensionless form.
After appropriate normalization they both are a function of one
dimensionless parameter only:

\begin{equation}
\hat{W}_{2x} = \hat{W}_{2y} =  F_2(N) = \ln{\left(1+\frac{1}{4
{N}^2}\right)}~.\label{powden2}
\end{equation}
Here $\hat{W}_{2x} = W_{2x}/W_{ox}^{(2)}$ and $\hat{W}_{2y} =
W_{2y}/W_{oy}^{(2)}$ are the normalized powers, while the
normalization constants $W_{ox}^{(2)}$ and $W_{oy}^{(2)}$ are
given by

\begin{equation}
\left(
\begin{array}{c}
{W}_{ox}^{(2)}\\ {W}_{oy}^{(2)}
\end{array}\right)= \left(
\begin{array}{c}
\mathcal{A}^2\\ \mathcal{B}^2
\end{array}\right)\frac{a_{2o}^2 I_o^2}{2 \pi c}~.
\label{W02}
\end{equation}
For practical purposes it is convenient to express Eq. (\ref{W02})
in the form:

\begin{equation}
\left(
\begin{array}{c}
{W}_{ox}^{(2)}\\ {W}_{oy}^{(2)}
\end{array}\right)= \left(
\begin{array}{c}
\mathcal{A}^2\\ \mathcal{B}^2
\end{array}\right) W_b \left[\frac{a_{2o}^2}{2 \pi}\right] \left[\frac{I_o}{\gamma
I_A}\right]~,\label{W02bis}
\end{equation}
\begin{figure}
\begin{center}
\includegraphics*[width=140mm]{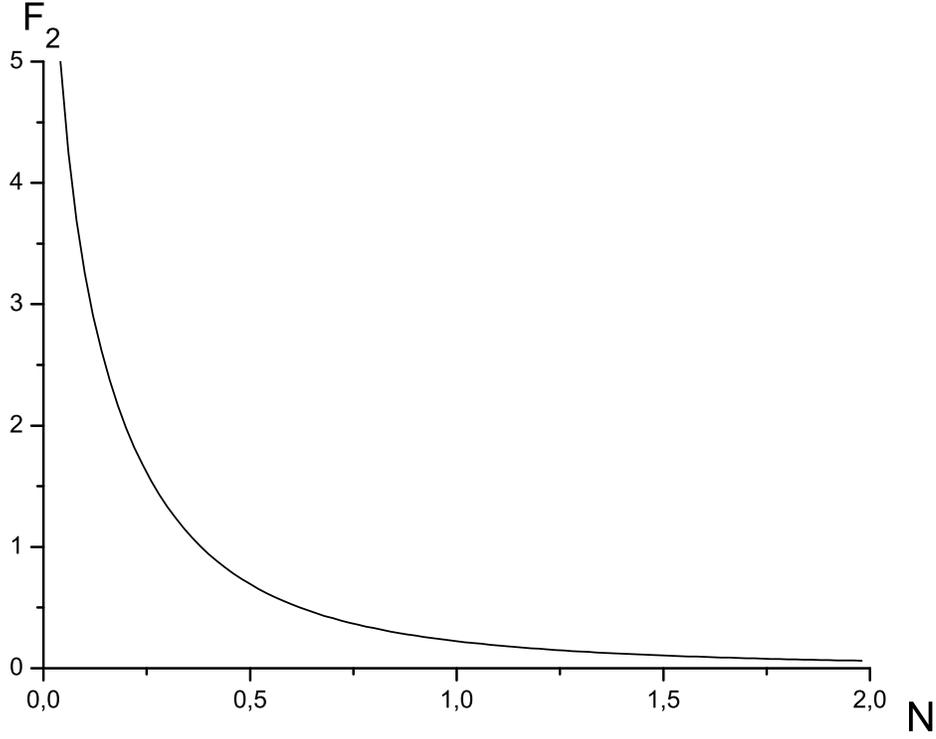}% Here is how to import EPS art
\caption{\label{W2} Illustration of the behavior of $F_2(N)$.}
\end{center}
\end{figure}
where $W_b = m_e c^2 \gamma I_o/e$ is the total power of the
electron beam and $I_A = m_e c^3/e \simeq 17$ kA is the Alfven
current.

The function $F_2(N)$ is plotted in Fig. \ref{W2}. The logarithmic
divergence in $F_2(N)$ in the limit for $N \ll 1$ imposes a limit
on the meaningful values of $N$. On the one hand, the
characteristic angle $\hat{\theta}_\mathrm{\max}$ associated with
the intensity distribution is given by
$\hat{\theta}_\mathrm{\max}^2 \sim 1/{N}$. On the other hand, the
expansion of the Bessel function in Eq. (\ref{undurad4}) is valid
only as $\hat{\theta}^2 \lesssim N_w$. As a result we find that
Eq. (\ref{powden2}) is valid only up to values of $N$ such that $N
\gtrsim N_w^{-1}$. However, in the case $N < N_w^{-1}$ we deal
with a situation when the dimensionless problem parameter $N$ is
smaller than the accuracy of the resonance approximation $\sim
N_w^{-1}$. In this situation our electrodynamic description does
not distinguish anymore between a beam with finite transverse size
and a point-like particle and,  for estimations, we should make
the substitution $\ln{(N)} \longrightarrow \ln{(N_w^{-1})}$.
%It should be noted that, as has been shown in
%\cite{METH}, the same condition applies in order for the electrons
%wiggling amplitude in the undulator to be smaller than the
%transverse size of the beam. In fact, indicating with $r_w =
%K/(k_w \gamma)$ the wiggling amplitude we have:
%
%\begin{equation}
%\frac{\sigma^2}{r_w^2} = \frac{2+K^2}{2K^2} \gamma_z^2 \sigma^2
%k_w^2 = \frac{2+K^2}{4K^2} (\pi N_w N) > 1~. \label{accur}
%\end{equation}
%%

We will now compare our results for the second harmonic with
already known results for the first. The case treated in
\cite{METH} corresponds to a modulation wavefront orthogonal to
the direction of propagation, exactly as specified here for the
second harmonic (i.e. perfect resonance with Eq. (\ref{expara})
and Eq. (\ref{exbot}) valid) and allows direct comparison of
results. The outcomes of \cite{METH} have been presented,
similarly to what has been done here for the second harmonic, in
dimensionless form. After appropriate normalization, one finds:

\begin{equation}
\hat{W}_{1}({N}) =  \frac{W_1}{W_o^{(1)}} = F_1(N) = \frac{2}{\pi}
\left[\arctan\left(\frac{1}{{N}}\right)+\frac{{N}}{2}
\ln\left(\frac{{N}^2}{{N}^2+1}\right) \right]~,\label{powden1}
\end{equation}
where the normalization factor $W_{o}^{(1)}$ is given by

\begin{equation}
W_{o}^{(1)}= W_b \left[2 \pi^2 a_{1o}^2\right]
\left[\frac{I_o}{\gamma I_A}\right] \left[\frac{K^2}{2+K^2}\right]
N_w A_{JJ}^2~,\label{W01}
\end{equation}
$A_{JJ}$ being given by

\begin{equation}
A_{JJ} = J_0\left(\frac{K^2}{4+2
K^2}\right)-J_1\left(\frac{K^2}{4+2 K^2}\right)~.\label{AJJdef}
\end{equation}
Here $a_{1o}$ is the analogous of $a_{2o}$ for the first harmonic.
For notational reasons, $a_{1o}$ is one half of the original
modulation level $a_\mathrm{in}$ in Eq. (27) of \cite{METH}. It
should also be noted that all ${N}$ in Eq. (\ref{powden1}) are
multiplied by a factor $1/2$ with respect to what is reported in
\cite{METH}. This is because we are referring all results to the
Fresnel number for the second harmonic.

\begin{figure}
\begin{center}
\includegraphics*[width=140mm]{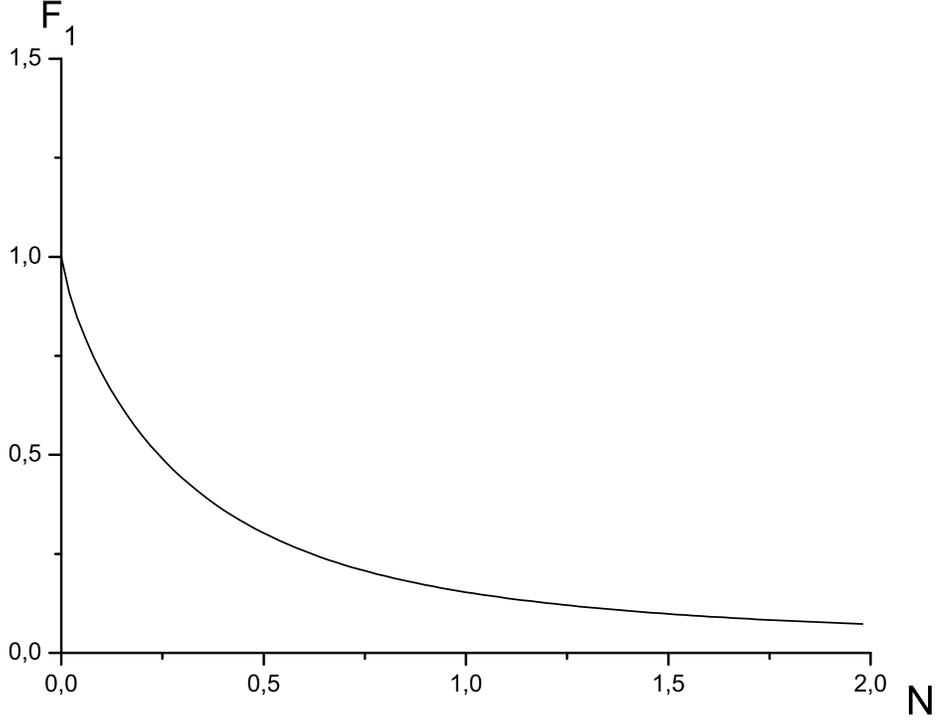}% Here is how to import EPS art
\caption{\label{F1} Illustration of the behavior of $F_1(N)$.}
\end{center}
\end{figure}
The function $F_1(N)$ is plotted in Fig. \ref{F1}. %Comparing Fig.
%\ref{W2} and Fig. \ref{F1} or, equivalently, Eq. (\ref{powden2})
%and Eq. (\ref{powden1}) we conclude that the first and the second
%harmonic normalized powers present the same behavior in $N$ in the
%limit  $N \gg 1$, but have different asymptotic as $N \ll 1$.
We can compare more quantitatively the normalized power for the
second and for the first harmonic:

\begin{eqnarray}
\frac{F_{2}(N)}{F_{1}(N)} = \frac{\pi}{2}
\left[\frac{\ln{\left(1+\frac{1}{4
{N}^2}\right)}}{\arctan{\left(\frac{1}{{N}}\right)}+\frac{{N}}{2}
\ln{\left(\frac{N^2}{N^2+1} \right)}}\right]~, \label{compx}
\end{eqnarray}
while, from a practical viewpoint, the comparison between the real
powers is equal to

\begin{equation}
\frac{W_2}{W_1} = \frac{W_{ox}^{(2)}+W_{oy}^{(2)}}{W_{o}^{(1)}}
\frac{F_{2}(N)}{F_{1}(N)} = \frac{1}{(2 \pi)^3 N_w}
\frac{2+K^2}{K^2}\frac{a_{2o}^2}{a_{1o}^2}
\frac{\mathcal{A}^2+\mathcal{B}^2}{A_{JJ}^2}
\frac{F_{2}(N)}{F_{1}(N)}~.\label{realcomp}
\end{equation}

\begin{figure}
\begin{center}
\includegraphics*[width=140mm]{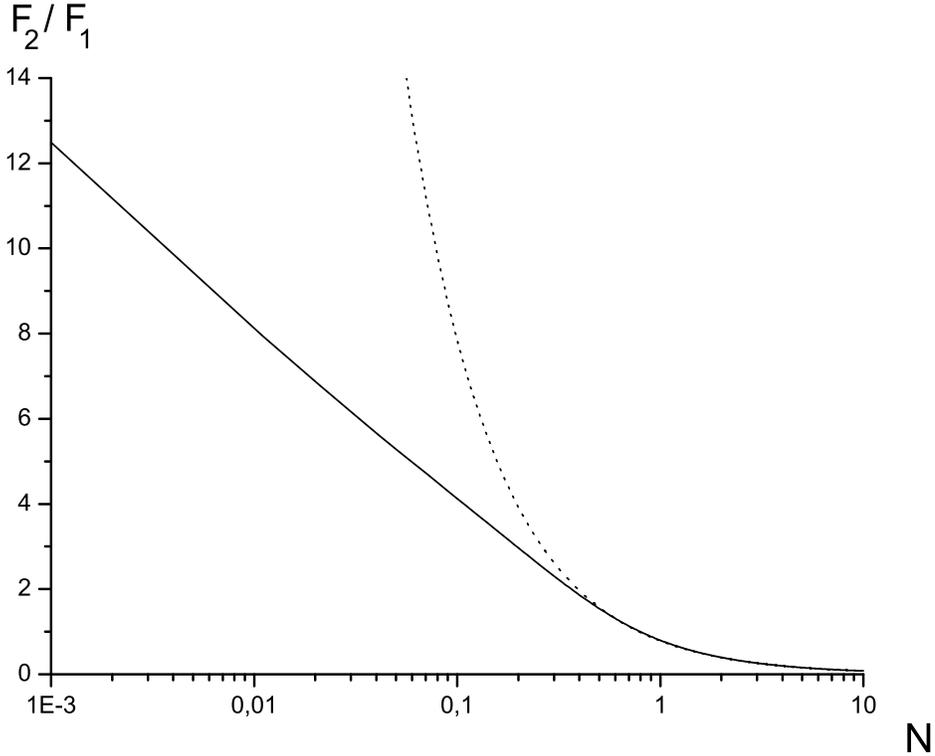}% Here is how to import EPS art
\caption{\label{unica} Solid line: illustration of the behavior of
$F_2/F_1$ as a function of ${N}$.  Dashed line: its asymptotic,
${\pi}/(4 N)$, for ${N}\gg 1$.}
\end{center}
\end{figure}
It is interesting to calculate Eq. (\ref{compx}) in the limit $N
\gg 1 $. We have

\begin{eqnarray}
\frac{F_{2}(N)}{F_{1}(N)} \longrightarrow
\frac{\pi}{4{N}}~~\mathrm{at} ~~ N\gg 1~. \label{compxlim}
\end{eqnarray}
In Fig. \ref{unica} we plot the behavior of $F_2/F_1$ as a
function of ${N}$ and its asymptotic, ${\pi}/{(4{N})}$, for $N \gg
1 $.

\begin{figure}
\begin{center}
\includegraphics*[width=140mm]{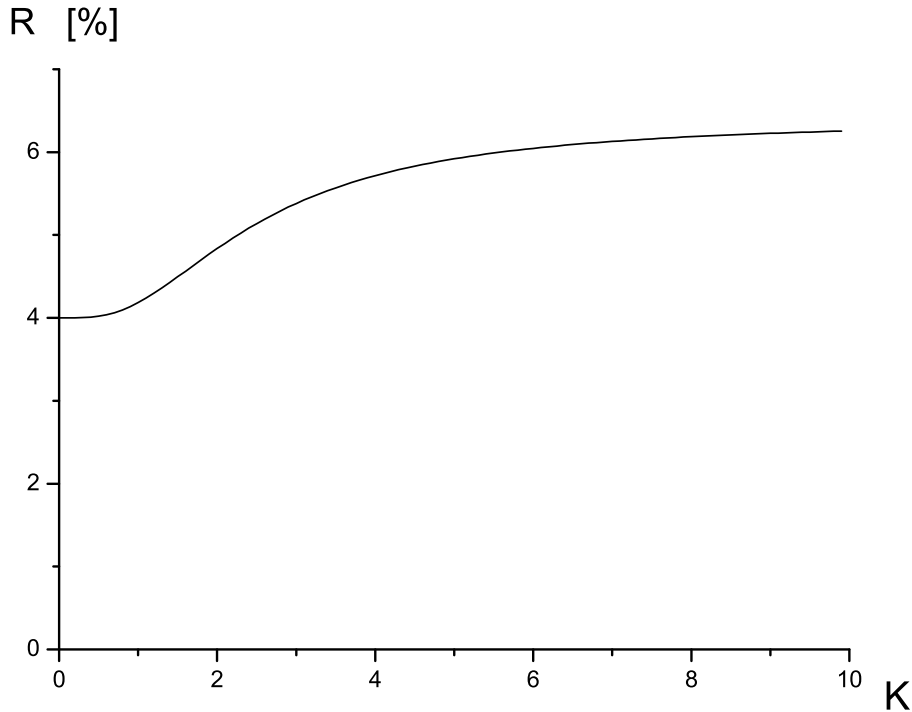}% Here is how to import EPS art
\caption{\label{R} Illustration of the behavior of the ratio
between the second harmonic power due to the $y$ vertical and the
$x$ horizontal polarization components, $R(K)$.}
\end{center}
\end{figure}
Finally, it is possible to study the ratio between the second
harmonic power due to the $y$ vertical and the $x$ horizontal
polarization components, that is only a function of the $K$
parameter and is simply given by the ratio $R = W_{2y}/W_{2x}$:

\begin{equation}
R(K) = \frac{W_{2y}}{W_{2x}} =
\frac{\mathcal{A}^2(K)}{\mathcal{B}^2(K)} ~.\label{WyWx}
\end{equation}
A plot of $R(K)$ is given in Fig. \ref{R}. As it is seen the
relative magnitude scales from $4 \%$ in the case $K \ll 1$ till
about $6 \%$ in the limit $K \gg 1$: as one can see, the vertical
polarization component of the radiation depends quite weakly on
the $K$ parameter. The knowledge of the polarization contents of
the radiation, even if relatively small as in this case, can be
important from an experimental viewpoint. For example, in the VUV
wavelength range, the reflection coefficients of many materials
(e.g. SiC, that is widely used for mirrors) exhibit a complicated
behavior, and there may be even an order of magnitude difference
depending on the polarization of the radiation. It should be noted
that $R(K)$ is independent of the particular model chosen for the
beam modulation as it is easy to understand inspecting Eq.
(\ref{undurad4}). It is also important to remark the fact that the
second harmonic radiation from a planar undulator is linearly
polarized, since vertical and horizontal polarization components
are characterized by the same phase factor. This fact is
well-known in Synchrotron Radiation theory for a single particle
and it is true for any observation angle and any harmonic of the
radiation from a planar undulator \cite{UNDU} in contrast, for
instance, to the case of bending magnet radiation, when vertical
and horizontal polarization components exhibit a relative $\pi/2$
phase shift, indicating circular polarization.

An important comment to what has been done before is needed. We
calculated the electric field, the angular intensity distribution
and the power for the second harmonic making a particular
assumption about the electron beam modulation in Eq.
(\ref{expara}). This amounts to consider the modulation wavefront
orthogonal to the direction of propagation of the beam. The same
assumption has been implicitly done calculating the first harmonic
power (the expression in \cite{METH} has been used, which does not
account for deflection angles). In this particular case we have
seen that the total power of the second harmonic radiation does
not depend on the deflection angles $\eta_x$ and $\eta_y$. In the
more general situation we find that the second harmonic power can
be independent of the beam deflection angle (like in the situation
treated by us) or can decrease due to the presence of extra
oscillating factors in $\vec{l}$ in Eq. (\ref{undurad5}). On the
contrary, in \cite{RHUA}, an increase of the total power is
reported due to the presence of deflection angles. %: such a
%conclusion is incorrect.

\section{\label{sec:stat} Discussion}

After the presentation of our theory has been given, in this
Section we want to discuss in a more detailed way the differences
between our approach and the currently accepted treatment of the
problem of second harmonic generation.

It is worth to begin summarizing the steps which led us to our
main results. First, we started from the wave equation assuming
that the electromagnetic sources are given externally by some code
calculating the electron beam bunching at the second harmonic.
Second, after applying the  ultrarelativistic approximation
($\gamma^2 \gg 1$) and the resonance approximation ($N_w \gg 1$),
both non-restrictive ones, we solved exactly the wave equation
using the Green's function method. Third, we calculated the
angular distribution of intensity assuming a given beam modulation
and we derived an expression for the total power radiated at the
second harmonic by integrating the expression for the angular
distribution of intensity. Finally we compared the expression for
the second harmonic power with the analogous expression for the
first harmonic.

In \cite{SCHM} and later on in \cite{KIM2, RHUA} the
ultrarelativistic and the resonance approximation were used too,
but several steps were performed on the wave equation which we
find incorrect. First, the gradient term in the source part of the
wave equation is overlooked. Second, the particles trajectory in
the transverse $x$ direction is expanded (we will comment on this
later on), and because of that the Fresnel number $N$ is not
identified as the main physical parameter of the problem. Third,
after these manipulations, the wave equation is not solved but,
rather, the second harmonic power is estimated in the following
way: (a) the squared of the (manipulated) source parts of the wave
equation for the second harmonic is calculated; (b) the squared of
the source parts for the first \cite{RHUA} or the third
\cite{SCHM, KIM2} harmonic is calculated; (c) the ratio between
the square of the source parts for the second and either the first
\cite{RHUA} or the third \cite{SCHM, KIM2} is taken. As a result,
magnitude of the second harmonic power and polarization
characteristics are predicted that are in disagreement with what
we have found. In \cite{RHUA}, further notions regarding the case
of deflection angle between the beam and the undulator axis are
introduced. We already expressed our critical view on this last
conclusion at the end of the previous Section. Let us briefly
comment on the other points mentioned above by analyzing more in
detail the approach followed in \cite{SCHM,KIM2,RHUA}.

As has been already said, we find that neglecting the gradient
term in the wave equation is not correct. As we have seen in
Section \ref{sec:ours} such term is responsible for a contribution
to the total intensity for the second harmonic both for the
horizontal and for the vertical polarization components and,
indeed, it cannot be neglected. Doing so would result in any case
in an overall incomplete result: namely one would obtain only part
of the horizontally polarized component of the field.

Going further with the derivation in \cite{SCHM,KIM2,RHUA}, the
motion of the electrons in the $x$-direction is written as a sum
of a fast oscillation due to the undulating motion and a slow
motion due to the betatron functions. On the $y$-direction
instead, only the slow motion due to betatron functions is
present. The beam distribution is then considered as a collection
of individual point-particles, i.e. a sum of $\delta$-Dirac
functions. For the $i$-th electron one may write

\begin{equation}
x'_i(z) = \bar{x}_i(z) + \Delta x_i(z)~ \label{assu1}
\end{equation}
and

\begin{equation}
y'_i(z) = \bar{y}_i(z)~, \label{assu2}
\end{equation}
where $\Delta x_i(z)$ describes the fast oscillation, while
$\bar{x}_i(z)$ and $\bar{y}_i(z)$ describe the slow motion.

All the $\delta$-Dirac in the $x$ coordinate on the right hand
side of Maxwell equation are subsequently expanded as

\begin{equation}
\delta(x_i-\bar{x}_i(z) - \Delta x_i(z)) \simeq \delta(x_i-
\bar{x}_i(z)) - \Delta x_i(z) \delta'(x_i-\bar{x}_i(z))~,
\label{expwrong}
\end{equation}
based on the only assumption that the transverse beam dimension is
much larger than the wiggling amplitude of the electron motion.
%\footnote{Note, by contrast the difference with our paper. We
%assumed from the very beginning, that the transverse size of the
%electron beam is "not smaller" (and not "much larger") than the
%typical wiggling motion of the electrons (compare also with Eq.
%(\ref{accur})) in order to justify Eq. (\ref{unp}). This
%assumption does not identify any physical parameter.}.
It should be noted that, based on this assumption, the ratio
between the wiggling amplitude and the transverse beam dimensions,
$K/(\gamma k_w \sigma)\ll 1$, is identified as the main physical
parameter of the theory and is denoted as the coupling strength of
the second harmonic emission. In contrast with this we have found
that the  exact solution of the wave equation depends on the
Fresnel number, but not on the coupling strength (see, for
instance, Eq. (\ref{I2xybis})). In this regard it is suggestive to
write $1/N = L_w c/(\sigma^2 \omega)$ as $1/N= [K/(\gamma k_w
\sigma)]^2\times \pi N_w (2+K^2)/(4 K^2) $. As $K/(\gamma k_w
\sigma)$ assumes a fixed value (for instance, much smaller than
unity according to the assumption above), our Fresnel number can
assume any value, depending on the number of undulator periods
$N_w$. If, on the one hand, $N \gg 1$ we have a behavior $F_2(N)
\sim 1/(4 N^2) \propto N_w^2$ so that $F_2(N)/F_1(N) \propto N_w$
as has been seen in Eq. (\ref{compxlim}) and therefore the ratio
$W_2/W_1$ is independent on $N_w$ as one can see from Eq.
(\ref{realcomp}). On the other hand, if $N \ll 1$ we obtain
$F_2(N) \sim \ln[1/(4N^2)] \sim \mathrm{const}$ so that
$F_2(N)/F_1(N)$ is independent on $N_w$ while $W_2/W_1 \propto
N_w^{-1}$. On the contrary, being based on the coupling strength
parameter only, current understandings of the second harmonic
mechanism predict that $W_2/W_1$ is always independent on $N_w$.

Finally we find that, from a mathematical viewpoint, the expansion
in Eq. (\ref{expwrong}) constitutes an incorrect manipulation of
the right hand side of Maxwell equation and. To show this, we
simply need to consider the mathematical structure of the wave
equation. For any polarization component we are dealing with a
differential equation:

\begin{equation}
\mathcal{L} \widetilde{{E}}_{x,y} (x,y,z) = f_{x,y}(z) \sum_i
\delta(x_i-\bar{x}_i(z) - \Delta x_i(z)) \delta (y_i -
\bar{y}_i(z))~, \label{diff}
\end{equation}
with

\begin{eqnarray}
\mathcal{L} = \left({\nabla_\bot}^2 + {2 i \omega \over{c}}
\frac{\partial}{\partial z}\right) ~,\label{mathquan}
\end{eqnarray}
where $f_{x,y}(z)$ is a function containing the appropriate phase
factor. This is essentially equivalent to our starting equation,
Eq. (\ref{incipit4}), the only differences being that the sources
are presented in a different way and that, at that stage, we  had
already assumed that the transverse beam dimensions are not
smaller than the wiggling amplitude of the electron motion.

The problem with the expansion in Eq. (\ref{expwrong}) is that
$\Delta x_i = \Delta x_i(z)$ is a function of the longitudinal
coordinate and that the Green's function for the wave equation
depends on both longitudinal and transverse coordinates. Let us
see this point in more detail. If we call with $G(z_o-z', x_o -
x', y_o -y')$ the Green's function of the operator $\mathcal{L}$
we have

\begin{eqnarray}
\tilde{E}_{x,y}(z_o,x_o,y_o) &=& \int_{-\infty}^{\infty} dx'
\int_{-\infty}^{\infty} dy'  \int_{-\infty}^{\infty} dz'
~G(z_o-z', x_o - x', y_o -y')\cr && \times f_{x,y}(z')
\delta(x'-\bar{x}(z') - \Delta x(z')) \delta (y' - \bar{y}(z)) ~,
\label{solE}
\end{eqnarray}
that is

\begin{eqnarray}
\tilde{E}_{x,y}(z_o,x_o,y_o) & = &\int_{-\infty}^{\infty} dz'
G(z_o-z', x_o -\bar{x}(z') - \Delta x(z'), y_o -\bar{y}(z'))\cr
&&\times f_{x,y}(z') ~. \label{solE2}
\end{eqnarray}
It follows that the expansion of the $\delta$-Dirac in Eq.
(\ref{expwrong}) is mathematically equivalent to the expansion of
the Green's function $G$ in $\Delta x_i(z')$ around
$x_o-\bar{x}_i(z')$; however  under the only assumptions
$\bar{x}_i(z') \gg \mid \Delta x_i(z') \mid$ and $\bar{y}_i(z')
\gg \mid \Delta x_i(z') \mid$ we cannot expand the Green's
function in $\Delta x_i(z')$ around $x_o - \bar{x}_i(z')$. In fact
we have that

\begin{eqnarray}
&&G(z_o-z', x_o-\bar{x}_i(z') - \Delta x_i(z'), y_o - \bar{y}(z'))
 \cr&& \ne G(z_o-z,x_o-\bar{x}_i(z'),y_o-\bar{y}_i(z'))\cr && -
\Delta x_i(z') \frac{d G(z_o-z',\xi,y_o-\bar{y}_i(z'))}{d
\xi}\mid_{\xi = x_o-\bar{x}_i(z')}~, \label{Gexpa}
\end{eqnarray}
because $G$ is simultaneously a function of $z$ not only through
$\Delta x_i (z')$, $\bar{x}_i(z')$ and $\bar{y}_i(z')$ but also
through $z_o-z'$.

\section{\label{sec:conc} Conclusions}

In this paper we addressed the mechanism of second harmonic
generation in Free-Electron Lasers.

We found that an early treatment of this phenomenon \cite{SCHM} is
based on arbitrary manipulations of the source term of the wave
equation, which describes the electrodynamical part of the
problem. First, an important part of the source term is neglected
and, second, an expansion of the particles trajectory in the
transverse horizontal direction is performed, while we find that
there is no ground for such a step. Moreover, this leads to the
identification of the ratio between the amplitude of the electron
wiggling motion in the (planar) undulator and the electron beam
transverse size as the main physical parameter, while such
parameter does not play any role in our theory. The same steps
were also followed in \cite{KIM2,RHUA}. After these manipulations,
the wave equation is not solved but, rather, an estimation of the
second harmonic power is given by calculating the squared of the
manipulated source parts in the wave equation for the second and
either for the first \cite{RHUA} or the third \cite{SCHM,KIM2}
harmonic and, subsequently, taking the ratio between the squared
of the second harmonic source part and either the squared of the
first or the third. Finally in \cite{RHUA} it is introduced the
notion of a second harmonic power increasing when a deflection
angle between the beam trajectory and the undulator $z$ direction
is present. On the contrary, we find that such power can only
decrease or, at most, be independent of the deflection angle,
depending on how the
beam modulation is prepared. %We conclude that \cite{KIM2,RHUA}
%predict a wrong dependence of the second harmonic power on the
%problem parameters and even of the polarization characteristics of
%the second harmonic field.

By solving analytically the wave equation with the help of the
Green's function technique we derived an exact expression for the
field of the second harmonic emission. We limited ourselves to the
steady-state case which is close to practice in High-Gain Harmonic
Generation (HGHG) schemes but, for the rest, we did not make
restrictive approximations. This solution of the wave equation may
therefore be used as a basis for the development of numerical
codes dealing with second harmonic emission, which should be using
as input data the electron beam bunching for the second harmonic
as calculated by self-consistent FEL codes.

We found that, in general, the second harmonic field presents both
horizontal and vertical polarization components and that the
electric field is linearly polarized, while the relative magnitude
of the power associated to the vertical polarization component to
that associated to the horizontal polarization component is a
function of the undulator deflection parameter only. Using our
result we calculated analytically the directivity diagram and the
power associated with the second harmonic radiation assuming a
particular beam modulation case. We expect that these expressions
may be useful for cross-checking of numerical results.

In this paper, we presented a theory of the second harmonic
generation mechanism in XFELs and pointed out several notions on
such mechanism which we consider incorrect. In this regard, it
should be noted that some of them appear to go beyond the subject
of harmonic generation itself. In fact we have seen that these
notions imply an increase of the second harmonic power when an
angle between the beam direction and the undulator axis is
present, as if this was a general property depending on Maxwell
equations. In other words, it looks like  the solution of Maxwell
equations for any electron beam in an undulator would yield a
non-trivial dependence on the angle between the trajectory and the
undulator axis. If so, this conclusion should be valid, in
particular, for a single particle as well. We find that this is
not correct: in fact, for a single particle in a undulator the
dependence of the electric field on the angle between the average
direction of the particle and the undulator axis, $\eta_{x,y}$, is
simply related to the chosen reference system, i.e., usually, one
with the $z$ axis aligned with the undulator axis. A simple
rotation of an angle $\eta_{x,y}$ to a system with the $z$ axis
aligned with the electron average velocity would give a result
independent on such angle. This means that the basic
characteristics of undulator radiation, in particular the
intensity distribution at fixed frequency and the spectrum at
fixed observation angle $\theta_{x,y}$, depend on the combination
$(\theta_{x,y}-\eta_{x,y})$ only. In other words the presence of
an angle $\eta_{x,y}$ between the electron direction and the
undulator axis has the only effect of introducing a rotation in
the expression of the electric field otherwise leaving unvaried
all its characteristics, including its resonance frequency.

\section{\label{sec:ackn} Acknowledgements}

The authors wish to thank Martin Dohlus (DESY) for useful
discussions and Josef Feldhaus (DESY) for his interest in this
work.

\end{document}